\documentclass[aps,pra,reprint,amsmath,amssymb,superscriptaddress,showpacs,nobibnotes]{revtex4-1}
\usepackage{CJK}
\usepackage{graphicx,SIunits}
\usepackage{comment}
\usepackage{bm}     
\usepackage{hyperref} 
\usepackage{dsfont}    
\usepackage{color}
\usepackage[normalem]{ulem}

\begin{document}
\begin{CJK*}{GB}{}

\title{Physical interpretation of nonlocal quantum correlation through local description of subsystems
}
\author{Tanumoy Pramanik}
\thanks{These authors contributed equally to this work}
\affiliation{State Key Laboratory for Mesoscopic Physics, School of Physics, Peking University, Beijing, 100871, China}
\affiliation{ Beijing Academy of Quantum Information Sciences, Beijing 100193, China}

\author{Xiaojiong Chen}
\thanks{These authors contributed equally to this work}
\affiliation{State Key Laboratory for Mesoscopic Physics, School of Physics, Peking University, Beijing, 100871, China}

\author{Yu Xiang}
\affiliation{State Key Laboratory for Mesoscopic Physics, School of Physics, Peking University, Beijing, 100871, China}

\author{Xudong Li}
\affiliation{State Key Laboratory for Mesoscopic Physics, School of Physics, Peking University, Beijing, 100871, China}
\author{Jun Mao}
\affiliation{State Key Laboratory for Mesoscopic Physics, School of Physics, Peking University, Beijing, 100871, China}
\author{Jueming Bao}
\affiliation{State Key Laboratory for Mesoscopic Physics, School of Physics, Peking University, Beijing, 100871, China}
\author{Yaohao Deng}
\affiliation{State Key Laboratory for Mesoscopic Physics, School of Physics, Peking University, Beijing, 100871, China}
\author{Tianxiang Dai}
\affiliation{State Key Laboratory for Mesoscopic Physics, School of Physics, Peking University, Beijing, 100871, China}

\author{Bo Tang}
\affiliation{Institute of Microelectronics, Chinese Academy of Sciences, Beijing 100029, China}
\author{Yan Yang}
\affiliation{Institute of Microelectronics, Chinese Academy of Sciences, Beijing 100029, China}
\author{Zhihua Li}
\affiliation{Institute of Microelectronics, Chinese Academy of Sciences, Beijing 100029, China}

\author{Qihuang Gong}
\affiliation{State Key Laboratory for Mesoscopic Physics, School of Physics, Peking University, Beijing, 100871, China}
\affiliation{ Beijing Academy of Quantum Information Sciences, Beijing 100193, China}
\affiliation{ Frontiers Science Center for Nano-optoelectronics \& Collaborative Innovation Center of Quantum Matter, Peking University, Beijing, 100871, China }
\affiliation{ Collaborative Innovation Center of Extreme Optics, Shanxi University, Taiyuan 030006, Shanxi, China}
\affiliation{ Peking University Yangtze Delta Institute of Optoelectronics, Nantong 226010, Jiangsu, China}
\author{Qiongyi He}
\affiliation{State Key Laboratory for Mesoscopic Physics, School of Physics, Peking University, Beijing, 100871, China}
\affiliation{ Beijing Academy of Quantum Information Sciences, Beijing 100193, China}
\affiliation{ Frontiers Science Center for Nano-optoelectronics \& Collaborative Innovation Center of Quantum Matter, Peking University, Beijing, 100871, China }
\affiliation{ Collaborative Innovation Center of Extreme Optics, Shanxi University, Taiyuan 030006, Shanxi, China}
\affiliation{ Peking University Yangtze Delta Institute of Optoelectronics, Nantong 226010, Jiangsu, China}
\author{Jianwei Wang}
\affiliation{State Key Laboratory for Mesoscopic Physics, School of Physics, Peking University, Beijing, 100871, China}
\affiliation{ Beijing Academy of Quantum Information Sciences, Beijing 100193, China}
\affiliation{ Frontiers Science Center for Nano-optoelectronics \& Collaborative Innovation Center of Quantum Matter, Peking University, Beijing, 100871, China }
\affiliation{ Collaborative Innovation Center of Extreme Optics, Shanxi University, Taiyuan 030006, Shanxi, China}
\affiliation{ Peking University Yangtze Delta Institute of Optoelectronics, Nantong 226010, Jiangsu, China}

\date{\today}

\begin{abstract}
\noindent Characterization and categorization of quantum correlations are both fundamentally and practically important in quantum information science. 
Although quantum correlations such as non-separability, steerability, and  non-locality can be characterized by different theoretical models in different scenarios with either known (trusted) or unknown (untrusted) knowledge of the associated systems,  such characterization sometimes lacks unambiguous to experimentalist.  In this work, we propose the physical interpretation of nonlocal quantum correlation between two systems.  In the absence of {\it complete local description} of one of the subsystems quantified by the {\it local uncertainty relation}, the correlation between subsystems becomes nonlocal. Remarkably, different nonlocal quantum correlations can be discriminated from a single uncertainty relation  derived under local hidden state (LHS)-LHS model only.
We experimentally characterize the two-qubit Werner state in different scenarios. 
\end{abstract}


\maketitle
 
\end{CJK*}

\section{Introduction}

Quantum correlation between two or more subsystems that cannot be described by local-causal theories is a key resource in quantum information science~\cite{EPR, Bell, Schro_1, Schro_2, Brunner, Horodecki, Jones07_1,Jones07_2, tele1,tele2,tele3,tele4,tele5,tele6, QKey1,QKey2, QKey3, QKey4, QCom1, QCom2}. 
A crucial task is to characterize, categorize and certificate different  quantum correlations. 
In general, quantum correlations can be described by the joint probability distribution of the events measured in the subsystems.  
For the bipartite quantum systems, the correlation is defined by  
\begin{eqnarray}
\mathcal{P}=\left\{P(a_{\mathcal{A}_i}, b_{\mathcal{B}_j}| \rho_{AB}) = \text{Tr}\left[\left(\Pi^{\mathcal{A}_i}_a\otimes\Pi^{\mathcal{B}_j}_b\right) \rho_{AB}\right]\right\}~~~
\label{Exp_Prob}
\end{eqnarray}
where $\rho_{AB}$ is the unknown state composed by  Alice's and Bob's  systems, and  $\Pi^{\mathcal{A}_i}_a$ ($\Pi^{\mathcal{B}_j}_b$) is the projective measurement having outcomes of $a$ ($b$) for the $\mathcal{A}_i$  ($\mathcal{B}_j$) observable. 
The characterization of correlation of the state $\rho_{AB}$ implies the measurement of the probability distribution $\mathcal{P}$. 
For example, to certify the Bell nonlocality, the distribution  $\mathcal{P}$ has to violate Bell inequalities~\cite{Bell, CHSH, Brunner}. 
Quantum correlations are further categorized by entanglement~\cite{Horodecki} and quantum steering~\cite{ Schro_1, Jones07_1, Ramanathan_2018}. 
Wiseman \textit{et. al.} proposed a framework to describe all the three quantum correlations for the bipartite system by considering three different scenarios having either known (trusted) or unknown (untrusted) knowledge of the system~\cite{Jones07_1,Jones07_2,CHRW_2011}: 
(1) $\rho_{AB}$ is entangled  if $\mathcal{P}$ can not be generated by a separable state having trusted measurement devices in both subsystems. 
(2) $\rho_{AB}$ is steerable if $\mathcal{P}$ can not be produced by a local hidden state (LHS) model, in the case that one subsystem owns trusted measurement device while the other remains untrusted. 
(3) $\rho_{AB}$ is Bell nonlocal if $\mathcal{P}$ is incompatible with the local hidden variable (LHV) interpretation 
and both measurement devices are untrusted. 
Categorizing quantum correlations regarding their capability of controlling measurement apparatuses have enabled important applications in quantum information, e.g.,  device independent (DI) or one-side DI quantum key distribution~\cite{DI_QKD1, MDI_QKD1, MDI_QKD2} and randomness generation~\cite{DI_Random}. 
We however note that non-separability, steerability, and Bell nonlocality can only be verified by the violations of their own inequalities, asking for a general framework of characterizing quantum correlations. 
The conceptual definition of known or unknown systems may also lead to confusion and ambiguity to experimentalist who usually can well control the system and measurement apparatuses.


\begin{figure*}[t]
\includegraphics[width=5.9in]{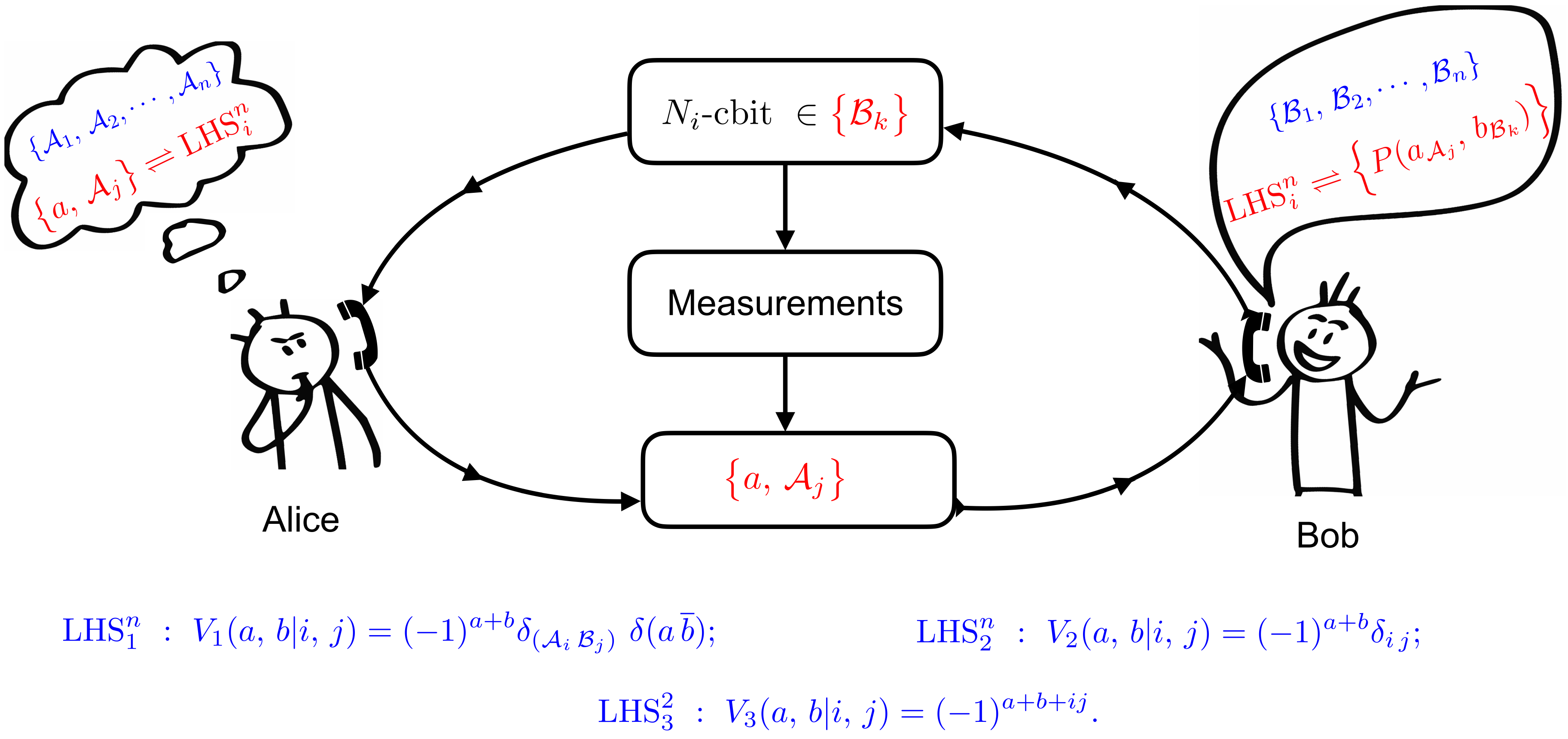}
\caption{Characterization of different nonlocal quantum correlations of the shared state $\rho_{AB}$ in a single local-description model. 
Bob's task is to characterize quantum correlations by the violations of the local-uncertainty relations. $\text{LHS}_i^n$, $i=1,2,3$, refers to different Bob's strategies in different scenarios to verify nonlocal correlations including entanglement, steerability and Bell nonlocality. $n$ is the number of measurement. 
Bob first asks Alice to minimize his uncertainty about the state of the system $B$ by communicating $N_i$-cbits (classical bits)  to Alice. 
Alice  then measures the appropriate observable $\mathcal{A}_j$ on the system $A$  and communicates the $\{a,\,\mathcal{A}_j\}$ information  back to Bob. 
Given the $\{a,\,\mathcal{A}_j\}$ information, Bob checks the uncertainty of the state of his system $B$. 
If the certain local-uncertainty relation is violated as Eq.(2) shown, Bob confirms that the shared state $\rho_{AB}$ is either entangled, steerable, or Bell nonlocal.  
}

\label{Fig_1}
\end{figure*}

\begin{figure}[b]
\includegraphics[width=3in]{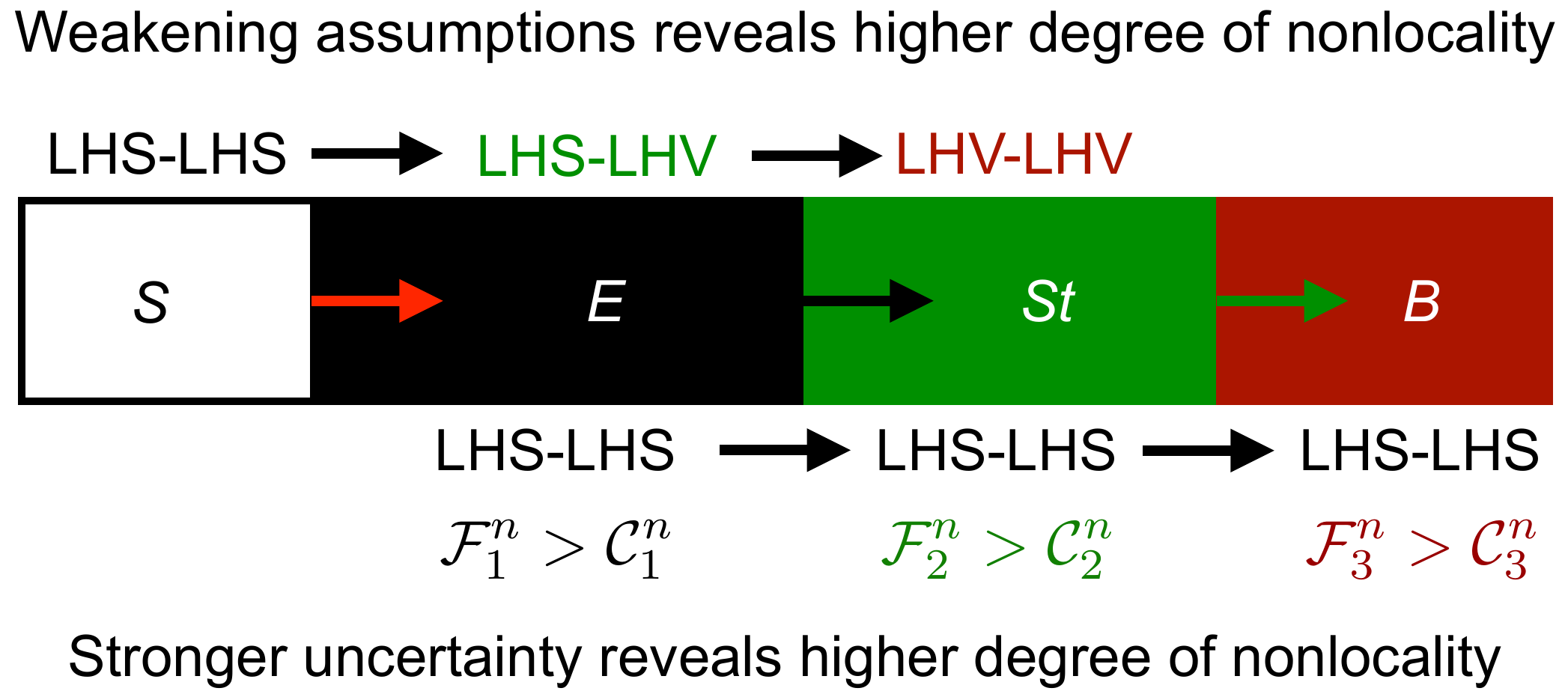}
\caption{ $S$, $E$, $St$, and $B$ correspond to separable, entangled, steerable, and Bell nonlocal correlation, respectively. Entanglement, steerability, and Bell nonlocal correlation are confirmed if the observed correlation $\mathcal{P}$ of Eq.~(\ref{Exp_Prob}) can not be explained by the theoretical models LHS-LHS, LHS-LHV, and LHV-LHV, respectively. Less assumption about the associated systems makes the correlation more nonlocal. In this work, we discriminate the degree of nonlocality under a single theoretical model, LHS-LHS with the help of proposed uncertainty relation of inequality~(\ref{G_UR}). Here, the violation of the inequalities $\mathcal{F}_1^n\leq \mathcal{C}_1^n$~(\ref{LHS_Ent}), $\mathcal{F}_2^n\leq \mathcal{C}_2^n$~(\ref{LHS_Seer}) and $\mathcal{F}_3^n\leq \mathcal{C}_3^n$~(\ref{LHS_CHSH}) validates entanglement, steerability and Bell nonlocal correlation, respectively.
}

\label{Fig_id}
\end{figure}

In this work, we propose a more physical interpretation of different nonlocal quantum correlations from {\it complete local description} of the subsystems that can be quantified by the {\it local uncertainty relation} of the subsystems. 
Our idea is inspired by the Einstein's comment~\cite{EPR} and Bell's seminal work~\cite{Bell} on incompleteness of quantum theory supplemented by LHV. 
We here ask a similar question:  {\it when two systems $A$ and $B$ are quantumly correlated, is there any complete local description of one of the subsystems, say, $B$ has nothing to do with $A$, or vice versa}? 
We will show how the local uncertainty relation {derived using the complete local description of subsystems} can help in discriminating different nonlocal quantum correlations. %
We remark that our way of characterizing  quantum correlations represents the fundamental connection of quantum nonlocality {and uncertainty relation.}
 Note that, in the previous works~\cite{Bell, CHSH, Brunner, Jones07_1,Jones07_2,CHRW_2011, Hofmann_2003, Zhen_2016}, the criteria of  on discrimination of different non-local quantum correlations are based on different forms of uncertainty relation formulated under LHS-LHS, LHS-LHV, LHV-LHV model. Here, we introduce single uncertainty relation~(inequality~(\ref{G_UR})) formulated under LHS-LHS model,  and this uncertainty relation can discriminate three different kinds of nonlocal correlations, e.g., entanglement, steering, Bell nonlocal correlation. See the Fig.~(\ref{Fig_id}) for more clear picture.
 
\section{Verification of different nonlocal correlations through complete description of subsystems}

Let us first consider the following game. 
Alice prepares a joint system of $A$ and $B$ in an unknown state $\rho_{AB}$, and sends the subsystem $B$ to Bob. While Bob may think that Alice can cheat him by preparing the state according to the LHS model, $\rho_{AB}^{\text{LHS}} = \sum_{i} p_i \rho^A_i\otimes \rho^B_i$, 
where $\rho^A_i$ ($\rho^B_i$) is Alice's (Bob's) local state, $p_i\geq 0$ and  $\sum_{i} p_i=1$. For $\rho_{AB}^{\text{LHS}}$, the system $B$ has complete local description, $\{p_i,\,\rho^B_i\}$. 
Here Bob tends to 
characterize the {nonlocal} quantum correlations of the  state $\rho_{AB}$ {with the help of} {the local uncertainty relation}. 
Bob asks Alice to minimize the uncertainty of the state of system $B$ by communicating $k$-cbit (classical bit) information. 
Given the $k$-cbit, Alice measures an observable of $\mathcal{A}_i$, and sends back the measurement outcome $a$ together with $\mathcal{A}_i$ to Bob. 
Finally, Bob checks {whether the joint probability distribution $\mathcal{P}$ can be describe by the complete local description of $\rho_{AB}^{\text{LHS}}$. This description is certified by } the uncertainty of their outcomes characterized by the condition $V(a,b|i,j)$ (where $i$, $j$ represent Alice's and Bob's choice of observables $\{\mathcal{A}_i\}$ and $\{\mathcal{B}_j\}$, {respectively}). 
Bob can confirm that the state $\rho_{AB}$ is entangled if  {the following local uncertainty} relation is violated 
\begin{eqnarray}
\mathcal{F}^n_k = \{\sum_{i,j=0}^{n-1} \sum_{a,b=0}^{1} V_k(a,b|i,j) P(a_{\mathcal{A}_i},\,b_{\mathcal{B}_j}| \rho_{AB}) \}  \leq \mathcal{C}^n_k, ~~~~~
\label{G_UR}
\end{eqnarray}
where  $V_k(a,b|i,j)$, $k\in\{1,2,3\}$, represents three different conditions of quantum correlations ($V_1$ for entanglement, $V_2$ for steering, and $V_3$ for Bell nonlocality); where  $n$ is the number of measurement performed on $A$ and $B$ and  chosen as   $n=2,3$ in our work (also in the experiment), which however can be chosen to an arbitrary number (see in the Appendix). 
The upper bound $\mathcal{C}^n_k$ is obtained by maximizing $\mathcal{F}^n_k$ over the state of $\rho_{AB}^{\text{LHS}}$ and Alice's all possible strategies. 
The violation of inequality~(\ref{G_UR}) implies that the shared state $\rho_{AB}$ cannot be written in the form of  $\rho_{AB}^{\text{LHS}}$. 

Figure~\ref{Fig_1} ~\cite{Fig1} sketches three different scenarios for the characterization and certification of quantum correlations in a single local-description model. 
 For simplicity, we start with the $LHS_2^n$ one, that is the LHS model description for quantum steering~\cite{Schro_2, Jones07_1,Jones07_2,Paul}).  

\subsection{Verification of steerability}

For the verification of steerability of the shared state $\rho_{AB}$, Bob asks to minimize his uncertainty of observables $\mathcal{B}_i$. He checks the uncertainty of their outcomes constrained by the condition of $V_2(a,b|i,j)=(-1)^{a+b} \delta_{ij}$, and the  {local uncertainty} relation thus turns into~\cite{Saunders, bennet12}
\begin{eqnarray}
\{\mathcal{F}_2^n=\sum_{i=0}^{n-1}|\langle \mathcal{A}_i\,\mathcal{B}_i\rangle| \} \leq \{C_2^n=\max_{\{\mathcal{A}_i\},\rho_{AB}^{\text{LHS}}}\left[\mathcal{F}_2^n\right]\},
\label{LHS_Seer}
\end{eqnarray}
{where the upper bound, $C_2^3 = \sqrt{3}$ ($C_2^2=\sqrt{2}$) for $n=3$ ($n=2$) measurement setting corresponds to the local description of Bob's system by the eigenstates of the observables $(\sigma_x\pm\sigma_y\pm\sigma_z)/\sqrt{3}$ ($(\sigma_x\pm\sigma_z)/\sqrt{2}$)} (see see in the Appendix for  details). 
The  $V_2$ shown as Eq.~(\ref{LHS_Seer}) represents Bob's residual uncertainty of the observable $\mathcal{B}_i$ (randomly chosen from a set of non-commuting observables~\cite{Joint_Measure_1,Joint_Measure_2,Joint_Measure_3}), given the $\{a,\,\mathcal{A}_i\}$ information from Alice. 
The classical communication of 1-cbit ($\log_2^{n=2}$) or 1.58-cbit ($\log_2^{n=3}$) is required from Bob to Alice when Bob randomly chooses $\mathcal{B}_i$ from a set of $n=2$  or 3 observables, say, $\{\sigma_z,\,\sigma_x\}$ or  $\{\sigma_z,\,\sigma_x,\,\sigma_y\}$, respectively. The violation of inequality~(\ref{LHS_Seer}) indicates that the system $B$ does not have complete {local description} independent of the system $A$, and the correlation is known as quantum {steering}~\cite{Saunders, bennet12, FgSt}. 


\subsection{Verification of Bell nonlocal correlation}

For the verification of Bell nonlocal, Bob does not reveal the choice of observables and there is no  communication from Bob to Alice. 
Given  information from Alice, Bob estimates the uncertainty from the measured  probability distribution $\mathcal{P}$. 
In the case of $n=2$ measurement, the uncertainty of input $\{i,j\}$ and output $\{a,b\}$ correlation is determined  by the CHSH game ({$V_3(a,b|i,j)=(-1)^{(a+b+ij)}$ which corresponds to the winning condition of 
the Clauser-Horne-Shimony-Holt game)~\cite{Bell,CHSH,Oppen}.}
Thus, the {local-uncertainty} Eq.~(\ref{G_UR}) can be rewritten as
\begin{eqnarray}
\mathcal{F}_3^2  \leq \{\max_{\{\mathcal{A}_i\},\,\rho_{AB}^{\text{LHS}}}\left[ \mathcal{F}_3^2\right]= 2\},
\label{LHS_CHSH}
\end{eqnarray}
where $\mathcal{F}_3^2=|\langle\mathcal{A}_0\left(\mathcal{B}_0+\mathcal{B}_1\right)\rangle + \langle\mathcal{A}_1\left(\mathcal{B}_0-\mathcal{B}_1\right)\rangle|$ {and the upper bound corresponds to local description of Bob's system by the state, e.g., $|0\rangle$}. 
The $V_3$  corresponds to Bob's residual uncertainty of the randomly chosen observables of $\{\mathcal{B}_0=\sigma_x,\,\mathcal{B}_1=\sigma_z\}$~\cite{Joint_Measure_1, Joint_Measure_4},  with respect to Alice's {individual} measurement from $\{\mathcal{A}_0, \mathcal{A}_1\}$. 
When the local uncertainty relation~(\ref{LHS_CHSH}) is violated, Bob validates the Bell nonlocal correlation~\cite{Bell,CHSH,Oppen}. 
The inequality~(\ref{LHS_CHSH}) becomes a necessary and sufficient condition for Bell nonlocality for the 2-measurement settings and binary outcomes~\cite{Horo_1995}.  
Note that the $\text{LHS}_3^2$ model is a stricter version of $\text{LHS}_2^n$ model, as the former represents a simultaneous steerability (uncertainty) of $\{\mathcal{B}_0,\,\mathcal{B}_1\}$~\cite{Oppen} while the later represents an {individual steerability (uncertainty) of $\mathcal{B}_i$ with respect to Alice's observable $\mathcal{A}_i$.} 
Therefore, the Bell nonlocal correlation becomes the strongest form of nonlocal correlations -- 
the violation of inequality~(\ref{LHS_CHSH}) indicates the violation of inequality~(\ref{LHS_Seer}). 

\subsection{Verification of entanglement}

To certify entanglement, Bob asks to minimize the value of $b$ for the $\mathcal{B}_j$ measurement, randomly chosen from the set of non-commuting observables. 
The classical communication of $2$-cbit (four possible combinations of $\{a,\mathcal{B}_j\}$) or $2.58$-cbit (six possible combinations) is required from Bob to Alice, when $n=2$ or 3-measurement is chosen, respectively. 
Bob evaluates the uncertainty of $V_1(a,b|i,j)=(-1)^{a+b} \delta_{(\mathcal{A}_i\mathcal{B}_j)} \delta_{(a\overline{b})}$, where $\overline{b}=b\oplus1$. 
Applying the condition of $V_1$ in the inequality~(\ref{G_UR}), the {local uncertainty} relation turns into
\begin{eqnarray}
\{\mathcal{F}_1^n &=& \sum_{i=0}^{n-1} P(\mathcal{A}_i,\,\mathcal{B}_i)\}
\leq  \{\max_{\rho_{AB}^{\text{LHS}}}\left[\mathcal{F}_1^n\right]= \mathcal{C}_1^n\}
\label{LHS_Ent}
\end{eqnarray}
where $P(\mathcal{A}_i,\,\mathcal{B}_i)=P(0_{\mathcal{A}_i},\,1_{\mathcal{B}_i}) + P(1_{\mathcal{A}_i},\,0_{\mathcal{B}_i})$, 0 and 1 refer to measurement outcomes; where $\mathcal{C}_1^{n=3}=2$, $\mathcal{C}_1^{n=2}=1$ {corresponds to Bob's local state, e.g., $|0\rangle$} (see details in the Appendix). 
The $V_1$  leads to the uncertainty of anti-correlated outcomes $a\oplus b=1$ when Alice and Bob both performs measurement of the same observable, i.e, $\mathcal{A}_i=\mathcal{B}_j$ on their respective subsystems. 
The violation of the {local uncertainty relation} in Eq.(\ref{LHS_Ent}) can confirm the presence of entanglement~\cite{Hofmann_2003, St_Ent3, St_Ent1, St_Ent2}. 
The uncertainty relation of Eq. (\ref{LHS_Ent}) is the weaker form of Eq.~(\ref{LHS_Seer}), as quantum steering considers uncertainty of all possible combinations of $\{a, b\}$, while entanglement only takes uncertainty of anti-correlated outcomes. 

\begin{figure}[]
\includegraphics[width=2.55in]{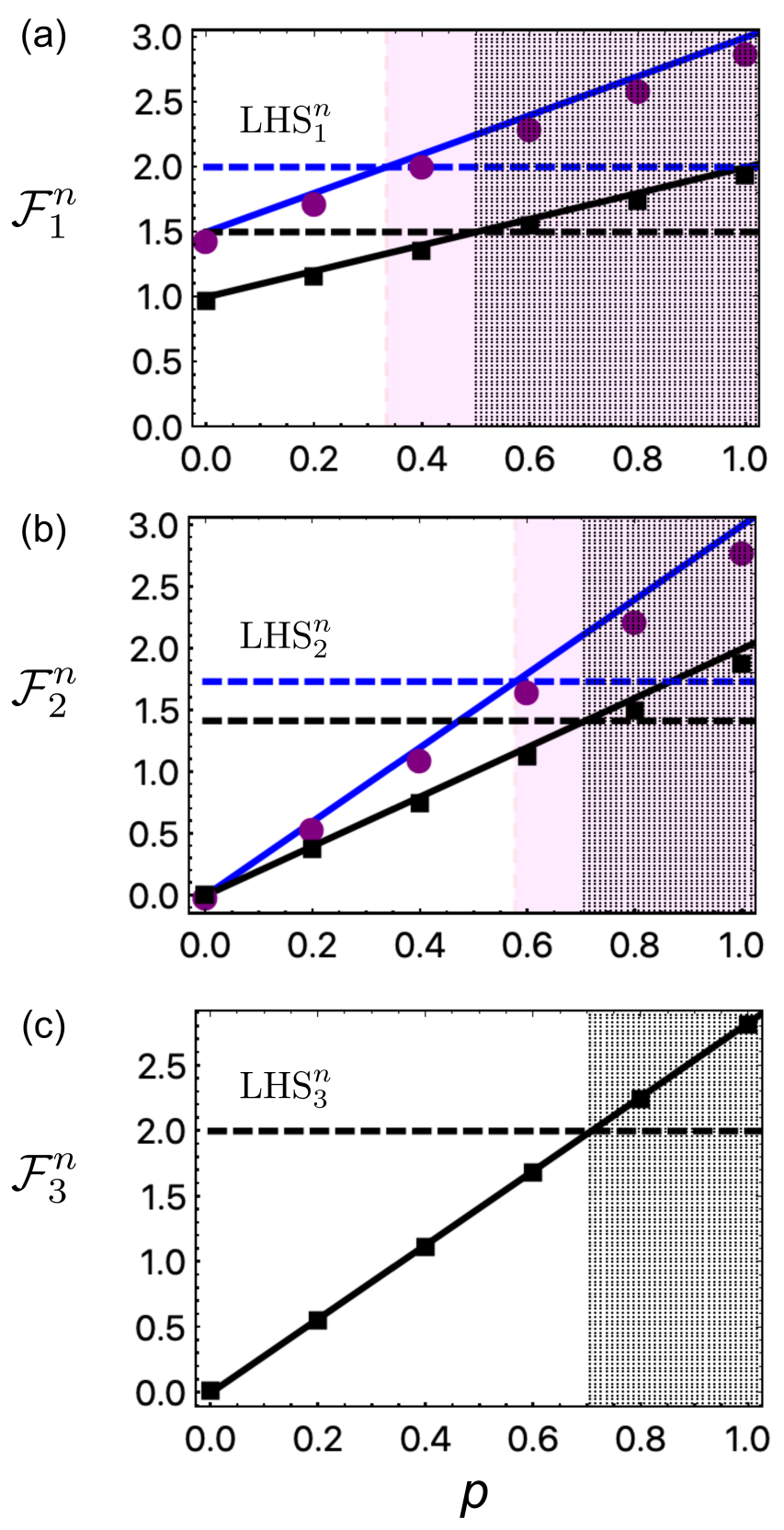}
\caption{Theoretical and experimental characterizations of (a) entanglement, (b) steerability, and (c) Bell nonlocality for the bipartite Werner state. 
The $\text{LHS}_i^n$, $i=1,2,3$ and $n=2,3$, refers to different LHS models in the three scenarios, see the derived local uncertainty relations of (3)--(5).  
All experiments were implemented on an integrated silicon-photonics quantum device. 
Points denote experimental data and lines denote theoretical prediction: circular and square points are for $n=3$ and $n=2$ measurement settings; blue and black lines are for $n=3$ and $n=2$ measurement, respectively.  
 Red shaded (black dotted) regime in (a)--(c) identifies the $p$ mixing parameter of the Werner state $\rho_W$, above which the  state is certified as entanglement, steerable, and Bell nonlocal, for $n=3$ ($n=2$) measurement settings, respectively. 
 Horizontal dashed lines are plotted for the guidance the achievable upper bound of the inequality value, $\mathcal{F}^n_k$.  
Note error bars ($\pm\sigma$) estimated from {20 sets of data}
are too small to be invisible in the plot.   
}
\label{Fig_2}
\end{figure}
\begin{figure}[t]
\includegraphics[width=3.3in]{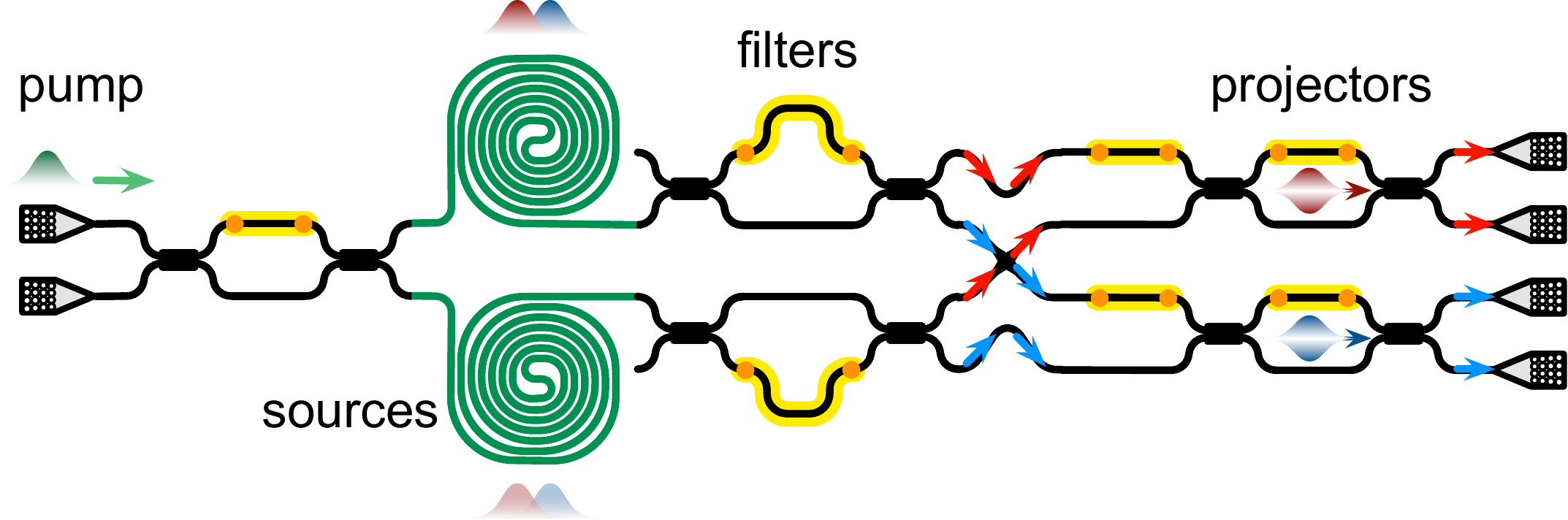}
\caption{{Schematic of an integrated silicon-photonics quantum device.} 
The quantum device is capable of generating, manipulating and analyzing  maximally path-entangled states. The device is fabricated on the silicon-on-insulator platform. Lines are silicon nanophotonic waveguides with the size of 450nm $\times$ 220nm, and yellow parts are  thermo-optic phase shifters that can be precisely controlled in experiment. 
A continuous wave laser light (at the wavelength of 1550.12nm) was used to pump two photon-pair sources, producing a pair of path-entangled photons via the spontaneous four-wave mixing (sFWM) process. 
The entangled photons were locally manipulated and analyzed by Alice (signal photon at 1545.31nm) and Bob (idler photon at 1554.91nm), respectively, which were implemented by the terminate Mach-Zehnder interferometers (MZIs).  The two photons are measured by two superconducting nanowire single-photon detectors (SNSPDs), and their coincidence were recorded by a time tagger. 
}
\label{FigScheme_text}
\end{figure}

\subsection{ Higher degree of nonlocality from more uncertainty of the condition $V_k(a,\,b|i,\,j)$}

For the purpose of connecting uncertainty relation~(\ref{G_UR}) with the degrees of nonlocality, we define normalized probability by $\mathcal{P}_k^n=\mathcal{F}_k^n/\max\left[\mathcal{F}_k^n\right]$, where $\max\left[\mathcal{F}_k^n\right]$ corresponds to algebraic maximum of $\mathcal{F}_k^n$. The corresponding uncertainty is measured by the Shannon entropy, $\mathcal{H}_k^n= - \mathcal{P}_k^n \log2\left[\mathcal{P}_k^n\right] - (1-\mathcal{P}_k^n) \log2\left[1-\mathcal{P}_k^n\right] $. Therefore, $\mathcal{H}_k^n$ determines the degree of uncertainty of the event $V_k(a,\,b|i,\,j)$, which corresponds to the correlation between Alice's and Bob's outcomes $a$ and $b$. For entanglement, steerability and Bell nonlocal correlation $\mathcal{H}_1^3 > 0.92$ (corresponds to the inequality~(\ref{LHS_Ent}), $\mathcal{F}_1^3 > 2$ and $\max\left[\mathcal{F}_1^3\right]=3$),  $\mathcal{H}_2^3 > 0.98$ (corresponds to the inequality~(\ref{LHS_Seer}), $\mathcal{F}_1^3 > \sqrt{3}$ and $\max\left[\mathcal{F}_2^3\right]=3$), $\mathcal{H}_3^2 > 1$ (corresponds to the inequality~(\ref{LHS_CHSH}), $\mathcal{F}_1^3 > 2$ and $\max\left[\mathcal{F}_3^2\right]=4$), receptively. As a result, higher degree of nonlocality implies larger threshold value of uncertainty of the condition $V_k(a,\,b|i,\,j)$.


Our local uncertainty relations as shown by inequalities of (3)--(5), which are all derived from a single inequality of (2) under different conditions, represent the more physical interpretation of different quantum correlations including quantum entanglement, steering and Bell nonlocal correlation. We now take the Werner state of $\rho_W= p\,\rho_{|\phi^-\rangle} + (1-p)\,\frac{\rm{I}\otimes \rm{I}}{4}$ as an example to test our local-description model in theory and experiment. 
 $\rho_{|\phi^-\rangle}$ is the density matrix of the singlet state of $|\phi^-\rangle=(|01\rangle - |10\rangle)/\sqrt{2}$, $I$ is the identity matrix, and $p$ ($0\leq p\leq 1$) denotes the mixing parameter.  
 The task now is to determine both in theory and experiment the bound of the $p$ parameter, above which the inequalities (3)--(5) can be violated and thus the state $\rho_W$ can be certified to be entangled, steerable, or Bell nonlocal. 
The results are shown in Fig.~\ref{Fig_2}.

\section{Experimental demonstration of different nonolocal correlations}

We experimentally verified the three quantum correlations for the Werner state. 
Figure \ref{FigScheme_text} shows the diagram of an integrated  silicon-photonic quantum device  that can generate, manipulate and analyze all four Bell states~\cite{Silverstone_15, Wang_16}. The integrated quantum device offers high levels of controllability and stabilities of operating quantum states of light~\cite{Wang:16D,Wang:review}. The {maximally entangled state has been } created with a high fidelity of $0.951\pm0.096$ by performing quantum state tomography (QST). 
The experimental realization of the $\rho_W$ state with a fully controllable mixture parameter $p$ is enabled by the classical mixture of quantum states (see experimental details in the Appendix).


Figure~\ref{Fig_2} shows the characterizations of entanglement, steering and Bell nonlocal,  experimentally 
demonstrating the violations of their inequalities of (\ref{LHS_Ent}), (\ref{LHS_Seer}) and (\ref{LHS_CHSH}), respectively. 
In Fig.~\ref{Fig_2}a, for $n=2$ and 3-measurement settings, entanglement is confirmed for $1/2<p\leq 1$ (black dotted) and $1/3<p\leq 1$ (red shaded), respectively~\cite{Hofmann_2003}. 
Note that 3-measurement is sufficient to fully reveal entanglement of the $\rho_W$ state up to the  value obtained by QST (see see in the Appendix and  Fig.~\ref{Fig_3}). 
In Fig.~\ref{Fig_2}b, quantum steerability is certified when $1/\sqrt{3}<p\leq1$ (red shaded) for the 3-measurement setting, larger than that for the 2-measurement setting having $1/\sqrt{2}<p\leq 1$ (black dotted). 
Increasing the number of measurement  of $n$ can relax the $p$ value of demonstrating steering~\cite{Saunders, bennet12}. For example, when implementing infinite measurement settings, the steerability inequality can be violated for $1/2<p\leq 1$~\cite{Jones07_2}. 
In Fig.~\ref{Fig_2}c, it shows that 
the state is demonstrated to be Bell nonlocal for $0.7071<p\leq 1$. Unlike the steering and entanglement scenarios, increasing the number of measurement to  three however does not relax the choice of $p$ parameter. 
 Bell nonlocality can be verified for $4/5<p\leq 1$ 
 using the $I_{3322}$ inequality, as reported in ref. ~\cite{BE3322}. 

Figure~\ref{Fig_3} summarizes the bound of violating the LHS inequalities for entanglement, steering and Bell nonlocal. We here consider the $\mathcal{F}^n_k$ for the $n=2,3$ measurement settings, and for infinite measurements and for QST measurement. The regimes of $p$ parameter obeying the LHS models are grayed, while the regimes for certificated entanglement, steerability, and Bell nonlocality are colored.  In the $\mathcal{F}^{465}_3$  bar, the red regime was estimated with 465 measurement settings~\cite{BNnm1}, and the black one refers to an unknown regime~\cite{AGT_06}.

\begin{figure}[ht]
\includegraphics[width=3.3in]{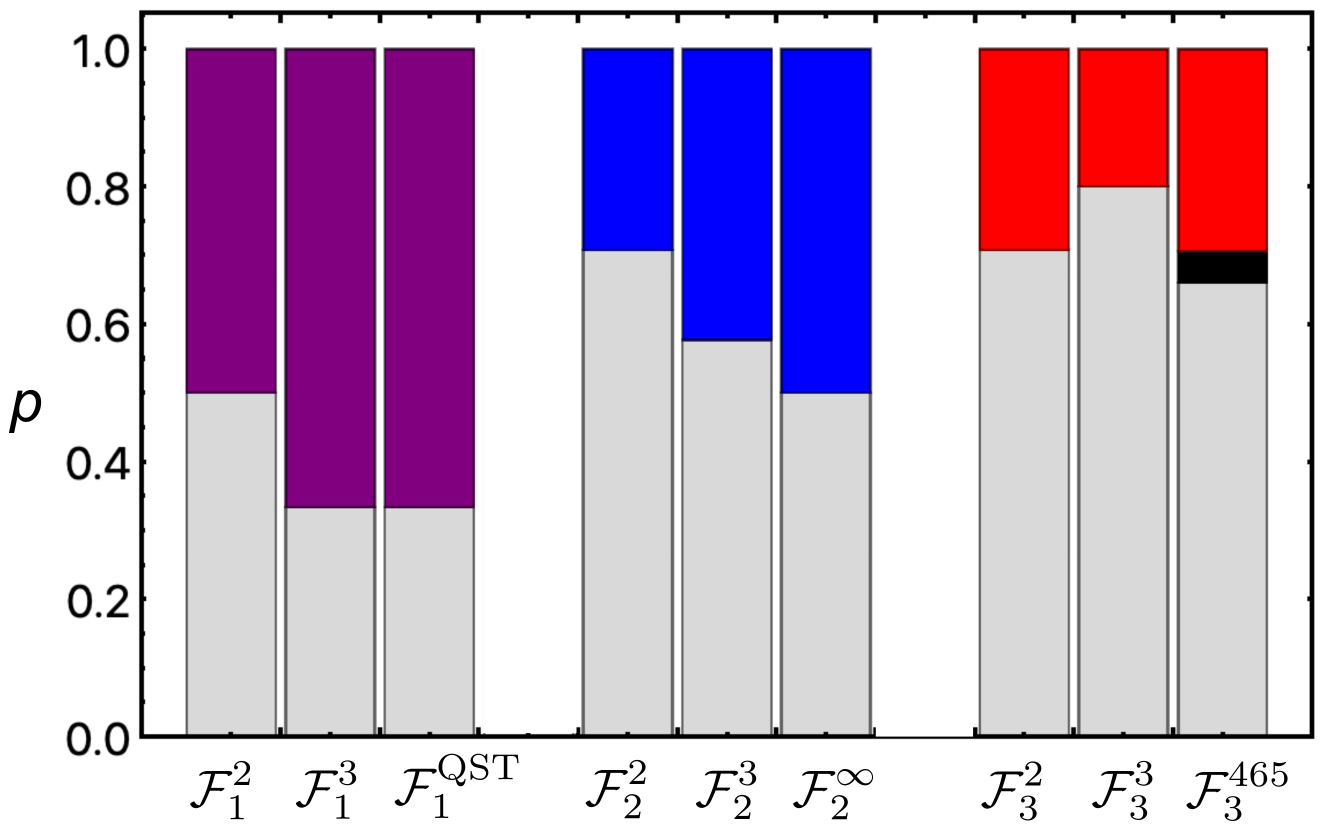}
\caption{Bound of inequality violation for the quantum correlations of entanglement ($\mathcal{F}^n_1$), steerability ($\mathcal{F}^n_2$) and Bell nonlocality ($\mathcal{F}^n_3$). 
The number of measurement settings $n=2,3$ are considered, while the $\mathcal{F}^{\infty}_k$ value is estimated from infinite measurement settings, and $\mathcal{F}^{\text{QST}}_k$ is quantified by QST. 
Purple, blue and red colored regime represents the bound of the $p$ mixture parameter, above which the state is certified as entanglement, steerable, and Bell nonlocal, respective. 
Grayed regimes denote the presence of LHS model. Note  the blacked regime refers to the inconclusive regime for Bell nonlocality. 
}
\label{Fig_3}
\end{figure}

 \section{Conclusions}

In sum,  we formulate single uncertainty relation under LHS-LHS model and different kinds of nonlocal correlations can be discriminated through it. This is major improvement over previously used different uncertainty relation based on different theortical models, e.g., LHS-LHS for entanglement, LHS-LHV for steering and LHV-LHV for Bell nonlocal correlation. We also show that different nonlocal quantum correlations have been characterized by the physical property, i.e,  complete local description of one of the subsystems, which is quantified by the local uncertainty relation {conditioned} on the outcomes of  subsystems. The violation of local uncertainty relation confirms the nonlocal correlation between subsystems. When increasing the uncertainty of the condition by restricting the communication between two parties, local uncertainty relation detects stronger form of nonlocal quantum correlation. Therefore, uncertainty of local description of one of the subsystems can be interpreted as nonlocal correlation between subsystems. 
As an example, in experiment, we have tested the uncertainty of local descriptions and the quantum correlation of subsystems prepared in the bipartite Werner states. The framework presented in this work may open new possibilities for interpretation of quantum correlation with respect to other fundamental properties of the multipartite systems.

\section{DATA AVAILABILITY}
The main data supporting the finding of this study are available within the article. Additional data can be provided upon request.

\section{ACKNOWLEDGMENTS}
We acknowledge support from  the Natural Science Foundation of China (nos 61975001, 61590933, 61904196, 61675007, 11975026), the National Key Research and development (R$\&$D) Program of China (2019YFA0308702, 2018YFB1107205,  2016YFA0301302), Beijing Natural Science Foundation (Z190005), and Key R$\&$D Program of Guangdong Province (2018B030329001). T.P. thanks Guruprasad Kar for useful discussion.\\

\section{AUTHOR CONTRIBUTIONS}

J.W. conceived the project. T.P., X.C., Y.X., X.L., J.M., J.B., Y.D., and T.D.  built the setup and carried out the experiment. Y.Y., B.T., and Z.L. fabricated the device. T.P., X.C., Y.X., X.L.,  J.W. and Q.H. performed the theoretical analysis. Q.H., Q.G., and J.W. managed the project. All authors discussed the results and contributed to the manuscript.

\section{COMPETING INTERESTS}
The authors declare no competing interests.

\section{ADDITIONAL INFORMATION}
Correspondence and requests for materials should be addressed to T.P. Emails to: tanu.pra99@gmail.com.

\onecolumngrid
\appendix

\section{Complete local description of quantum  system}
\label{Apdx_Complete}

In the seminal work~\cite{EPR}, Einstein et al. discussed about incompleteness of quantum theory in the presence of entanglement. J. Bell extended this work by considering quantum theory supplemented by the  hidden variable, say, $\lambda$ distributed according to $\{P(\lambda)\}$ where the restriction on $\lambda$  are $P(\lambda) \geq 0$ and $\sum_{\lambda} P(\lambda)=1$~\cite{Bell}. They showed that in the presence of entanglement between the system $A$ and $B$, 
\begin{eqnarray}
|\psi\rangle_{AB} = \frac{|01\rangle - |10\rangle}{\sqrt{2}},
\label{Singlet_State}
\end{eqnarray}
quantum theory can be incomplete~\cite{EPR, Bell}. The incompleteness occurring from the violation of Bell inequality by the correlation $\mathcal{P}$ of Eq.~(\ref{Exp_Prob_SM}) is known as Bell nonlocality, and the corresponding correlation is known as Bell nonlocal correlation. There are other well-known nonlocal quantum correlations, e.g., entanglement and steering. These nonlocal correlations have been explained with respect to trusting-untrusting scenarios~\cite{Jones07_1,Jones07_2}.

In this work, we have revisited the idea introduced by Einstein~\cite{EPR} et al. and Bell~\cite{Bell} considering the following question : when the systems $A$ and $B$ are quantumly correlated, is there any {\it complete local description} of the subsystems? Interestingly, if two systems $A$ and $B$ are in the separable state of
\begin{eqnarray}
\rho_{AB}^{\text{LHS}} = \sum_{i} p_i \rho^A_i\otimes \rho^B_i,
\label{LHS_State_SM}
\end{eqnarray}
where $\rho^A_i$ ($\rho^B_i$) is Alice's (Bob's) local state, $p_i\geq 0$ , and  $\sum_{i} p_i=1$, individual system $A$ ($B$) has {\it complete local description},  $\{p_i,\, \rho^{B}_i\}$ ($\{p_i,\, \rho^{A}_i\}$), i.e., systems $A$ and $B$ do not share quantum correlation. For the verification of the nonlocal correlation of the bipartite state $\rho_{AB}$, let us consider the following game. In this game, Alice prepares two quantum systems $A$ and $B$ in the an entangled state $\rho_{AB}$, and sends the system $B$ to Bob. Bob thinks that Alice may cheat him by preparing the system $A$ and $B$ in the separable state of the form of Eq.~(\ref{LHS_State_SM}). For the verification of nonlocal correlation of the shared state $\rho_{AB}$, Bob asks Alice to reduce his uncertainty about the state of the system $B$ for the measurement of observables chosen randmonly from the set of non-commuting observables $\{\mathcal{B}_j\}$. Therefore, communication of $k$-cbit (classical-bit) is required from Bob to Alice. According to Bob's information, Alice measures $\mathcal{A}_i$ on her system and communicates the measurement outcome $a$ and the choice of observables $\mathcal{A}_i$. From the information $\{a,\,\mathcal{A}_i\}$, Bob constructs the joint probability distribution 
\begin{eqnarray}
\mathcal{P}=\left\{P(a_{\mathcal{A}_i}, b_{\mathcal{B}_j}; \rho_{AB}) = Tr[\left(\Pi^{\mathcal{A}_i}_a\otimes\Pi^{\mathcal{B}_j}_b\right).\rho_{AB}]\right\}
\label{Exp_Prob_SM}
\end{eqnarray}
and checks the uncertainty of the {\it condition} characterized by $V_k(a,b|i,j)$. $V_k(a,b|i,j)$ corresponds to  the desired correlation between outcomes $a$ and $b$ for the measurement of observables $\mathcal{A}_i$ and $\mathcal{B}_j$, and $k\in{1,\,2,\,3}$ corresponds to three different conditions. The {\it local uncertainty} relation associated with the condition $V_k(a,b|i,j)$ becomes
\begin{eqnarray}
\mathcal{F}_k^n = \sum_{i,j=0}^{n-1}\sum_{a,b=0}^1 V_k(a,b|i,j) P(a_{\mathcal{A}_i,b_{\mathcal{B}_j}}|\rho_{AB}) \leq \mathcal{C}_k^n,
\label{G_UR_SM}
\end{eqnarray}
where $n$ is the number of observables chosen by Alice and Bob, $k\in\{1,\,2,\,3\}$ corresponds three different condition for discrimination of three different nonlocal correlations (i.e., entanglement, steering and Bell nonlocal correlation). The upper bound $\mathcal{C}_k^n$ is obtained by maximizing $\mathcal{F}_k^n$ over the shared state $\rho_{AB}^{\text{LHS}}$ and Alice's all possible strategies. The violation of the inequality~(\ref{G_UR_SM})
implies that the shared state $\rho_{AB}$ can not be written in the form of $\rho_{AB}^{\text{LHS}}$ of Eq.~(\ref{LHS_State_SM})  and Bob system does not have {\it complete local description} under considered $V_k(a,b|i,j)$. As a result, Bob validates the nonlocal correlation of the shared state $\rho_{AB}$.

\section{Incomplete local-description of the system $B$ due to the presence of entanglement}
\label{Apdx_Ent}

\begin{figure}[t]
\includegraphics[width=4in]{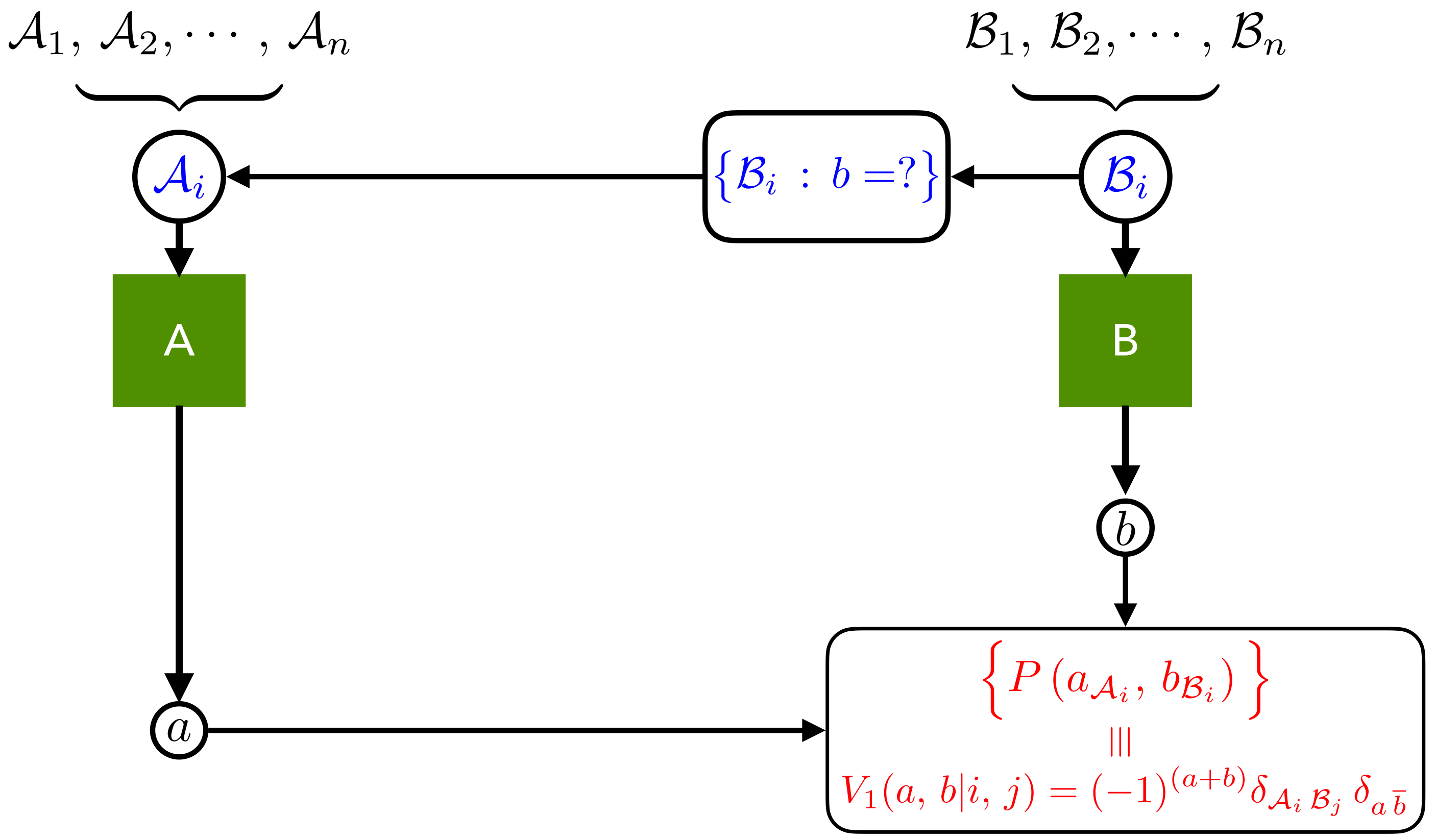}
\caption{ Bob's strategy to verify whether the experimentally observed correlation $\{P(a_{\mathcal{A}_i},\,b_{\mathcal{B}_i})\}$ infers the complete local description of the system $B$. Entanglement is confirmed when $\{P(a_{\mathcal{A}_i},\,b_{\mathcal{B}_i})\}$ cannot be explained by the condition $V_1(a,\,b|i,\,j) = (-1)^{a+b} \delta_{\mathcal{A}_i,\,\mathcal{B}_j}\,\delta_{a\,\overline{b}}$.
}
\label{Fig_Ent_SM}
\end{figure}

{\it $LHS_1^n$ : } Fig.~(\ref{Fig_Ent_SM}) describes the schematic diagram to certify the entanglement of the shared state $\rho_{AB}$. For the verification of entanglement of the shared state $\rho_{AB}$, Bob checks the uncertainty of the condition
\begin{eqnarray}
V_1(a,\,b|i,\,j) = (-1)^{a+b} \delta_{\mathcal{A}_i,\,\mathcal{B}_j}\,\delta_{a\,\overline{b}},
\label{V_Ent_SM}
\end{eqnarray}
where $\delta_{\mathcal{A}_i,\,\mathcal{B}_j}=1$ and $\delta_{a\,\overline{b}}=1$ only for $\mathcal{A}_i=\mathcal{B}_j$ and $a=\overline{b}=b\oplus 1$, respectively. In simple words, Bob checks the uncertainty of anti-correlation of their outcomes, $a\oplus b=1$ when they measure same observable $\mathcal{A}_i=\mathcal{B}_j$ on their respective systems. As a result, Bob's needs to send the information of $\{b,\,\mathcal{B}_j\}$ to Alice, and it requires 2-cbit ($\log_2^{k=4}$, corresponding to the four different combinations of two different outcomes and two measurement settings) or 2.58-cbit ($\log_2^{k=6}$,  corresponding to the six different combinations of two different outcomes and three measurement settings) for the choice of two measurement settings or three measurement seetings, respectively. In the case of verification condition, $V_1(a,\,b|i,\,j)$ of Eq.~(\ref{V_Ent_SM}), the inequality~(\ref{G_UR_SM}) becomes
\begin{eqnarray}
\mathcal{F}_1^n &=& \sum_{i=0}^{n-1} P(0_{\mathcal{A}_i},\,1_{\mathcal{B}_i}) + P(1_{\mathcal{A}_i},\,0_{\mathcal{B}_i})
\leq  \max_{\rho_{Ab}^{\text{LHS}}}\left[\mathcal{F}_1^n\right]=  \mathcal{C}_1^n
\label{LHS_Ent_SM}
\end{eqnarray}
where maximization is taken over all possible choices of $\rho_{AB}^{\text{LHS}}$. For the choice of two measurement settings, say, $\mathcal{A}_1=\mathcal{B}_1=\sigma_x$ and $\mathcal{A}_2=\mathcal{B}_2=\sigma_x$, the upper bound of the inequality~(\ref{LHS_Ent_SM}) becomes $\mathcal{C}_1^{n=2}=1$. In the case of three measurement settings, $\mathcal{A}_1=\mathcal{B}_1=\sigma_x$,  $\mathcal{A}_2=\mathcal{B}_2=\sigma_y$, $\mathcal{A}_3=\mathcal{B}_3=\sigma_z$, $\mathcal{C}_1^{n=3} =2$. In this scenario, the optimal cheating strategy for Alice corresponds to $\mathcal{F}_1^n=\mathcal{C}_1^n$ and the system $B$ has complete local description $\in\{|0\rangle, |1\rangle, \sqrt{\alpha} |0\rangle + \sqrt{1-\alpha}|1\rangle\}$. Note here that the above complete local description is not unique. The inequality~(\ref{LHS_Ent_SM}) is the {\it local uncertainty relation}, where `{\it local}' signifies that uncertainty relation is satisfied by the quantum systems having {\it complete local description} as shown in the Eq.~(\ref{LHS_State_SM}). From the violation of the uncertainty relation~(\ref{LHS_Ent_SM}), Bob validates the nonlocal correlation between system $B$ and $A$ and corresponding nonlocal correlation called entanglement~\cite{St_Ent1, St_Ent2, St_Ent3}.   The violation of the inequality~(\ref{LHS_Ent_SM}) is the necessary criterion for verification of entanglement. 
Alice's knowledge about Bob's strategy makes the criterion~(\ref{LHS_Ent_SM}) weakest. As a result, entanglement is the weakest nonlocal correlation.

\section{Incomplete local description of the system $B$ due to the presence of steerability}
\label{Apdx_Steer}

\begin{figure}[t]
\includegraphics[width=4in]{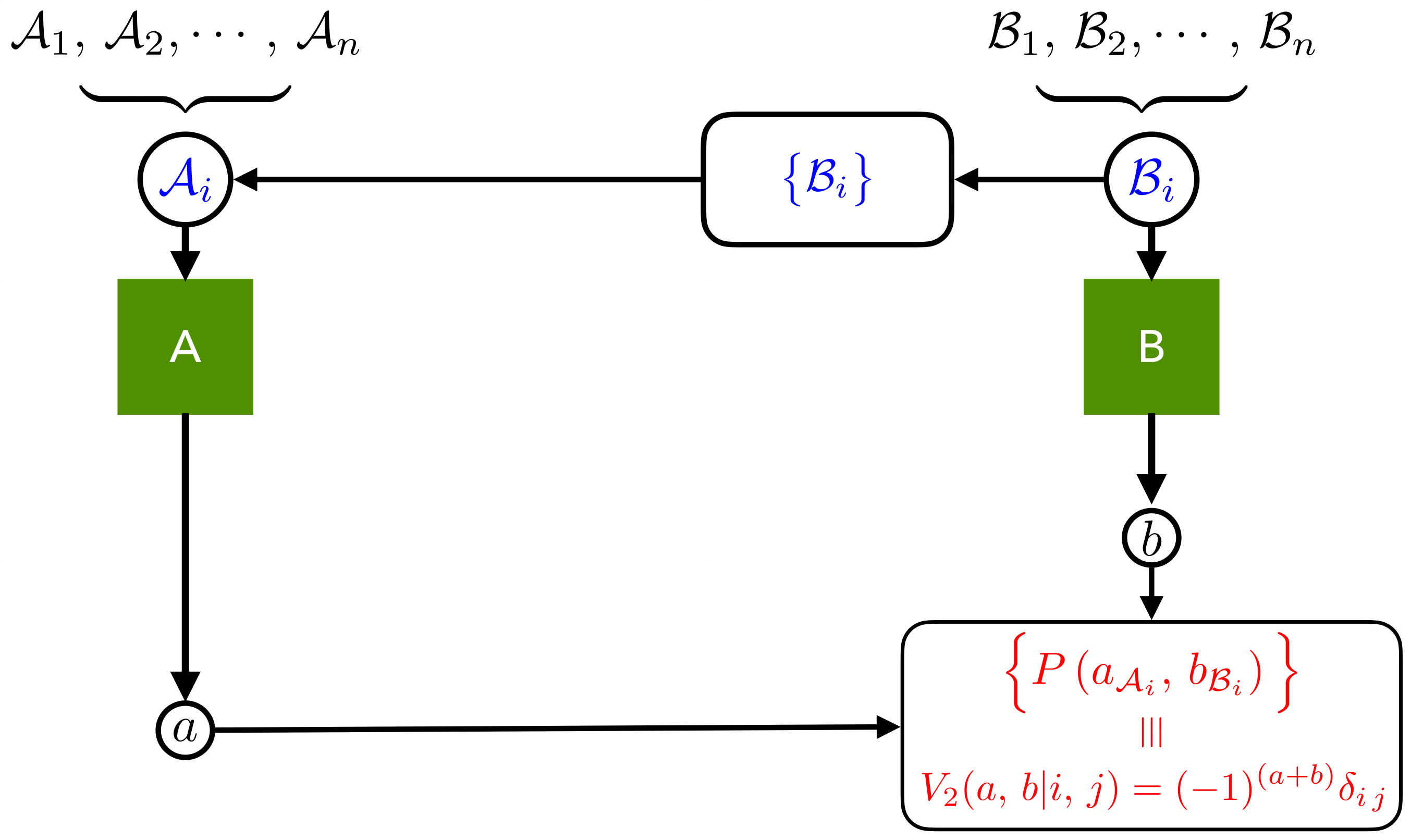}
\caption{ Bob's strategy to verify whether the experimentally observed correlation $\{P(a_{\mathcal{A}_i},\,b_{\mathcal{B}_i})\}$ corresponds to the complete local description of the system $B$. Steerability is certified when $\{P(a_{\mathcal{A}_i},\,b_{\mathcal{B}_i})\}$ cannot be explained by the condition $V_2(a,\,b|i,\,j) = (-1)^{a+b} \delta_{i,\, j}$.
}
\label{Fig_Steer_SM}
\end{figure}

{\it $LHS_2^n$ : }  Fig.~(\ref{Fig_Steer_SM}) describes the scenario to verify the steerability of the shared state $\rho_{AB}$. To verify steerability of the shared state $\rho_{AB}$, Bob checks the uncertainty of the condition
\begin{eqnarray}
V_2(a,\,b|i,\,j) = (-1)^{a+b} \delta_{i,\, j},
\label{V_St_SM}
\end{eqnarray}
which corresponds to Bob's residual uncertainty associated with the measurement of non-commuting observables $\{B_i\}$ from the knowledge of $\{a,\,\mathcal{A}_i\}$. For consideration of $V_2(a,\,b|i,\,j)$ of Eq.~(\ref{V_St_SM}), the uncertainty relation~(\ref{G_UR_SM}) becomes 
\begin{eqnarray}
\mathcal{F}_2^n=\sum_{i=0}^{n-1}|\langle \mathcal{A}
_i\,\mathcal{B}_i\rangle| \leq C_2^n=\max_{\{\mathcal{A}_i\},\rho_{AB}^{\text{LHS}}}\left[\mathcal{F}_2^n\right] ,
\label{LHS_Steer_SM}
\end{eqnarray}
where maximization is taken over all possible choices of $\{\mathcal{A}_0,\,\mathcal{A}_1,\, \cdots, \mathcal{A}_{n-1}\}$ and $\rho_{AB}^{\text{LHS}}$. When Bob randomly choose observable from the  set of two (three) non-commuting observables, say, $\{\sigma_x, \,\sigma_z\}$ ($\{\sigma_x, \,\sigma_y,\,\sigma_z\}$) the upper bound $C_2 = \sqrt{2}$ ($C_3=\sqrt{3}$) occurs when Bob's system $B$ has a complete local description by the eigenstates  of observables $\frac{\sigma_x\pm\sigma_z}{\sqrt{2}}$ ($\frac{\sigma_x\pm\sigma_y\pm\sigma_z}{\sqrt{3}}$). The violation of the inequality~(\ref{LHS_Steer_SM}) implies nonlocal correlation between Bob's system and Alice's system, and it is known as steering~\cite{Saunders, bennet12, FgSt}. Note here that the uncertainty relation~(\ref{LHS_Ent_SM}) is weaker form of the uncerainty relation~(\ref{LHS_Steer_SM}) as inequality~(\ref{LHS_Ent_SM}) deals with the uncertainty of $a\oplus b =1$ while inequality~(\ref{LHS_Steer_SM}) corresponds to the uncertainty of both the condition $a\oplus b =1$ and $a\oplus b =0$. 
Therefore, the violation of uncertainty relation~(\ref{LHS_Steer_SM}) indicates a stronger nonlocal correlation than the nonlocal correlation characterized by the uncertainty relation~(\ref{LHS_Ent_SM}). Using $\langle\mathcal{A}_i\mathcal{B}_i\rangle = Ps\left(\mathcal{A}_i,\,\mathcal{B}_i\right) - Pd\left(\mathcal{A}_i,\,\mathcal{B}_i\right)$ (where $Ps\left(\mathcal{A}_i,\,\mathcal{B}_i\right)= P(0_{\mathcal{A}_i},\,0_{\mathcal{B}_i}) + P(1_{\mathcal{A}_i},\,1_{\mathcal{B}_i})$) and $Pd\left(\mathcal{A}_i,\,\mathcal{B}_i\right) + Ps\left(\mathcal{A}_i,\,\mathcal{B}_i\right)=1$, inequality~\ref{LHS_Steer_SM} becomes 
\begin{eqnarray}
|2\sum_{i=1}^{n} Pd\left(\mathcal{A}_i,\,\mathcal{B}_i\right) - n|  \leq C_n.
\label{LHS_Steer_2_SM}
\end{eqnarray}
Violation of the above inequality indicate the violation of the inequality~(\ref{LHS_Ent_SM}), but the reverse is not true. Therefore, all steerable states are entangled and steerblity is stronger form of nonlocal correlation than entanglement.

Note that the violation of inequality~(\ref{LHS_Steer_SM}) is the necessary criterion for steerability, and it becomes more efficient to capture steerability of the given state with increment of the number of measurement settings, $n$~\cite{Saunders, bennet12}. But if inequality~(\ref{LHS_Steer_SM}) is satisfied, unsteerability can not be concluded.  
The unsteerability of the given state can be verified from the sufficient criteria~\cite{USt1}, but it  does not tell about steerability.

\section{ Incomplete local-description of the system $B$  due to the presence of Bell nonloclality}
\label{Apdx_Bell}

\begin{figure}[t]
\includegraphics[width=4in]{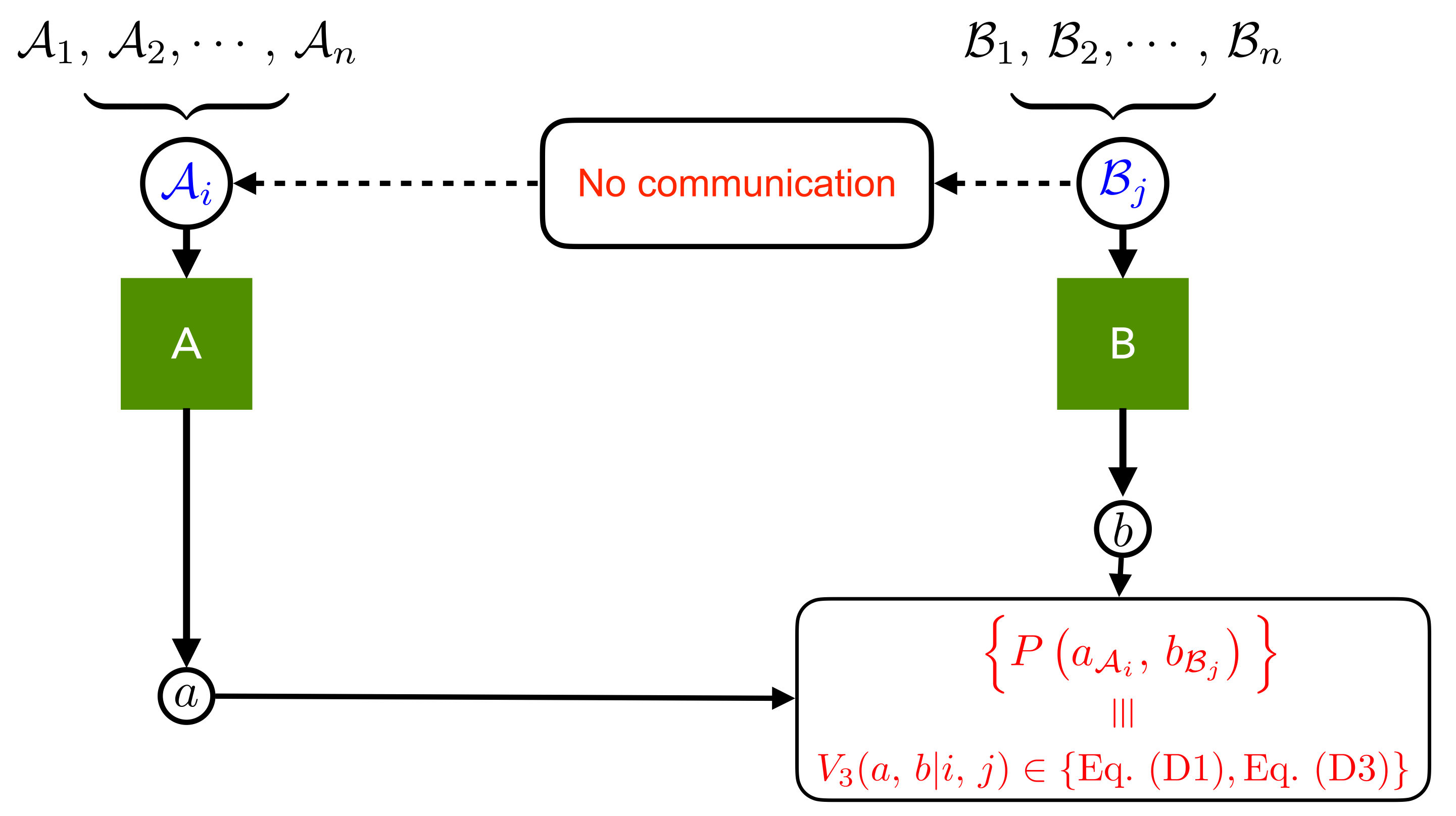}
\caption{ Bob's strategy to verify whether the experimentally observed correlation $\{P(a_{\mathcal{A}_i},\,b_{\mathcal{B}_i})\}$ corresponds to the complete local description of the system $B$. Bell nonlocality is verified when $\{P(a_{\mathcal{A}_i},\,b_{\mathcal{B}_i})\}$ cannot be explained by the condition $V_3(a,\,b|i,\,j)\in\{\text{Eq.(\ref{V_BI2_SM})},\,\text{Eq.(\ref{V_BI3_SM})}\}$.
}
\label{Fig_BI_SM}
\end{figure}

{\it  $LHS_3^n$ : } Fig.~(\ref{Fig_BI_SM}) describes the scenario to validate the Bell nonlocal correlation of the shared state $\rho_{AB}$. In the case of Bell nonlocality, Bob keeps secret the information of the observable $\mathcal{B}_i$ going to be measured on the system $B$. In the case of two measurement settings, he checks the uncertainty of the condition
\begin{eqnarray}
V_3(a,\,b|i,\,j) = (-1)^{(a+b+ij)},
\label{V_BI2_SM}
\end{eqnarray}
which corresponds the winning condition of CHSH game~\cite{Bell,CHSH,Oppen}. The uncertainty relation~(\ref{G_UR_SM}) for the choice of $V_2(a,\,b|i,\,j)$ of Eq.~(\ref{V_BI2_SM}) becomes
\begin{eqnarray}
 \mathcal{F}_3^2  \leq \max_{\{\mathcal{A}_i\},\,\rho_{AB}^{\text{LHS}}}\left[ \mathcal{F}_3^2\right]= 2,
\label{LHS_BI_2_SM}
\end{eqnarray}
where $\mathcal{F}_3^2=|\langle\mathcal{A}_0\left(\mathcal{B}_0+\mathcal{B}_1\right)\rangle + \langle\mathcal{A}_1\left(\mathcal{B}_0-\mathcal{B}_1\right)\rangle|$ corresponds to Bob's residual uncertainty of the  set of non-commuting observables $\{\mathcal{B}_0,\,\mathcal{B}_1\}$ from the knowledge of individual $\{a,\,\mathcal{A}_i\}$ and $i\in{0,\,1}$.  The above inequality is the necessary and sufficient criterion for Bell nonlocality for 2-measurement-setting and 2-outcome scenario~\cite{Horo_1995}. The equality $BI_2=2$ occurs when Bob's system has a complete local description by $\{|0\rangle,\,|1\rangle,\, \sqrt{\alpha} |0\rangle+ \sqrt{1-\alpha}|1\rangle\}$.  Note that the above local description is also not unique.

\begin{figure}[b]
\includegraphics[width=4in]{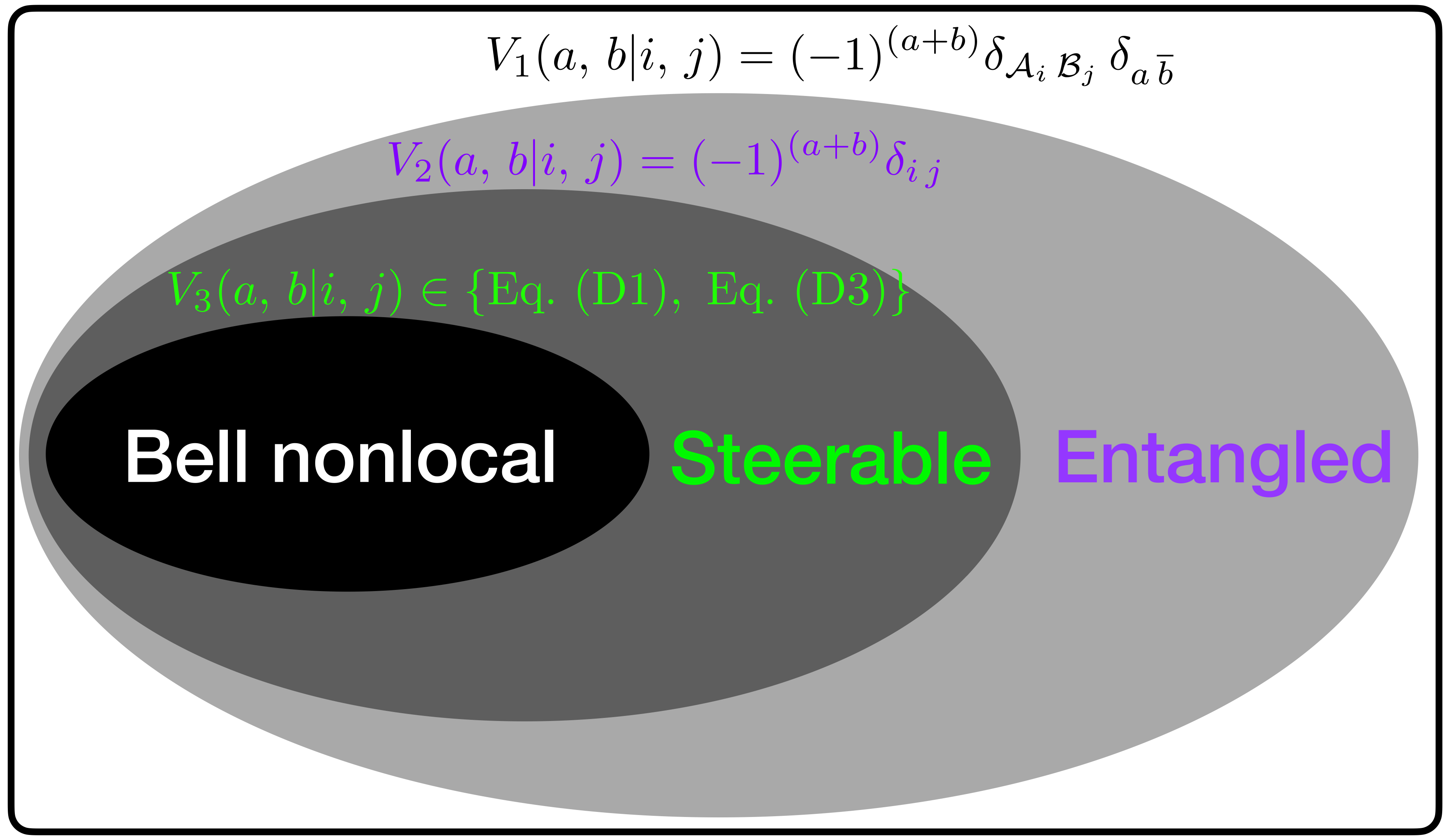}
\caption{ Different nonlocal quantum correlations. Bell nonlocal correlation is the strongest form of all nonlocal correlations, and entanglement is the weakest while steering lies between Bell nonlocal correlation and entanglement.
}
\label{Fig_TC}
\end{figure}

Recent developments improve Bell inequalities with $n$ number of measurement settings per side~\cite{BNnm1, BE3322, BI31, BI33}. Among them Bell inequality given in the Ref.~\cite{BE3322} is the inequivalent class of Bell-CHSH inequality and it can detect those entangled state which is Bell-CHSH local. Particularly, for the Werner state, Bell inequality with $465$ measurement settings per side has been developed~\cite{BNnm1} which can show Bell nolocality of Werner state for $0.7056<p\leq 1$. This inequality~\cite{BNnm1} gives the tightest Bell violation for Werner state after Bell-CHSH inequality. Therefore, Bell-CHSH inequality~(\ref{LHS_BI_2_SM}) works better than the existing 3-setting Bell inequality. Here, for three measurement settings, we consider Bell inequality given in the Ref.~\cite{BE3322} as an example.  Bob checks the following condition
\begin{eqnarray}
V_2(a,\,b|i,\,j) =(-1)^{a\oplus b=\max[0,i.j-1]} (1-\delta_{i,3}\delta_{j,3}) + (-1)^a\,(1-\delta_{i,3}) - (-1)^b (1-\delta_{j,3}).
\label{V_BI3_SM}
\end{eqnarray}
Under this condition, the inequality~(\ref{G_UR_SM}) becomes
\begin{eqnarray}
\mathcal{F}_3^3   \leq \max_{\{\mathcal{A}_i\},\,\rho_{AB}^{\text{LHS}}}\left[ \mathcal{F}_3^2\right]= 4,
\label{LHS_BI_3_SM}
\end{eqnarray}
where $\mathcal{F}_3^3= | \langle\mathcal{A}_1\left(\rm{I} + \mathcal{B}_1\rangle + \mathcal{B}_2 + \mathcal{B}_3\rangle \right)  +\langle\mathcal{A}_2\left(\rm{I}+\mathcal{B}_1 +\mathcal{B}_2 - \mathcal{B}_3\right)\rangle + \langle\mathcal{A}_3\left(\mathcal{B}_1 -\mathcal{B}_2\right)\rangle - \langle\mathcal{B}_1\rangle - \langle\mathcal{B}_2\rangle |$, and Bob chooses observables randomly from the set $\{\mathcal{B}_1=\sigma_z,\,\mathcal{B}_2=\sin(\frac{\pi}{3})\sigma_x+\cos(\frac{\pi}{3})\sigma_z,\,\mathcal{B}_3=\sin(\frac{2\pi}{3})\sigma_x+\cos(\frac{2\pi}{3})\sigma_z\}$.
In this scenario, $|0\rangle$ becomes one of the complete local description of the system $B$ corresponding to $\mathcal{F}_3^3=4$. The inequality~(\ref{LHS_BI_3_SM}) detects Bell nonlocality of the Werner state~(\ref{W_State_SM}) for $p>0.8$, whereas the inequlality~(\ref{LHS_BI_2_SM}) confirms Bell nonlocality for $p>1/\sqrt{2}$.

Bell-CHSH inequality~(\ref{LHS_BI_2_SM}) works better for the two-qubit Werner state than existing Bell inequalities with three measurement settings and two outcomes scenario~\cite{BE3322, BNnm1, BI31, BI33}.  Note that, mathematically, inequality~(\ref{LHS_BI_2_SM}) can be written as
\begin{eqnarray}
BI_2= |\langle \mathcal{A}_1^\prime \mathcal{B}_1\rangle + \langle \mathcal{A}_2^\prime \mathcal{B}_2\rangle| \leq \sqrt{2},
\end{eqnarray}
where $\mathcal{A}_1=\frac{\mathcal{A}_1+\mathcal{A}_2}{\sqrt{2}}$ and $\mathcal{A}_2=\frac{\mathcal{A}_1-\mathcal{A}_2}{\sqrt{2}}$. Therefore, violation of inequality~(\ref{LHS_BI_2_SM}) implies the violation of inequality~(\ref{LHS_Steer_SM}) while the reverse is not true. Therefore, all Bell nonlocal states are steerable. Unlike steerability,  Bell nonlocality includes the uncertainty of choice of Bob's observables from the set $\{\mathcal{B}_1,\,\mathcal{B}_2, \cdots, \mathcal{B}_n\}$ which makes Bell nonlocal correlation as a strongest form nonlocal correlation than steerability and entanglement. A complete picture of all the correlations can be viewed as in Fig.~(\ref{Fig_TC}), which represents that Bell nonlocal correlation forms a subset of both steering and entanglement, while steering is itself a subset of entanglement.

\begin{figure}[b]
\includegraphics[width=4in]{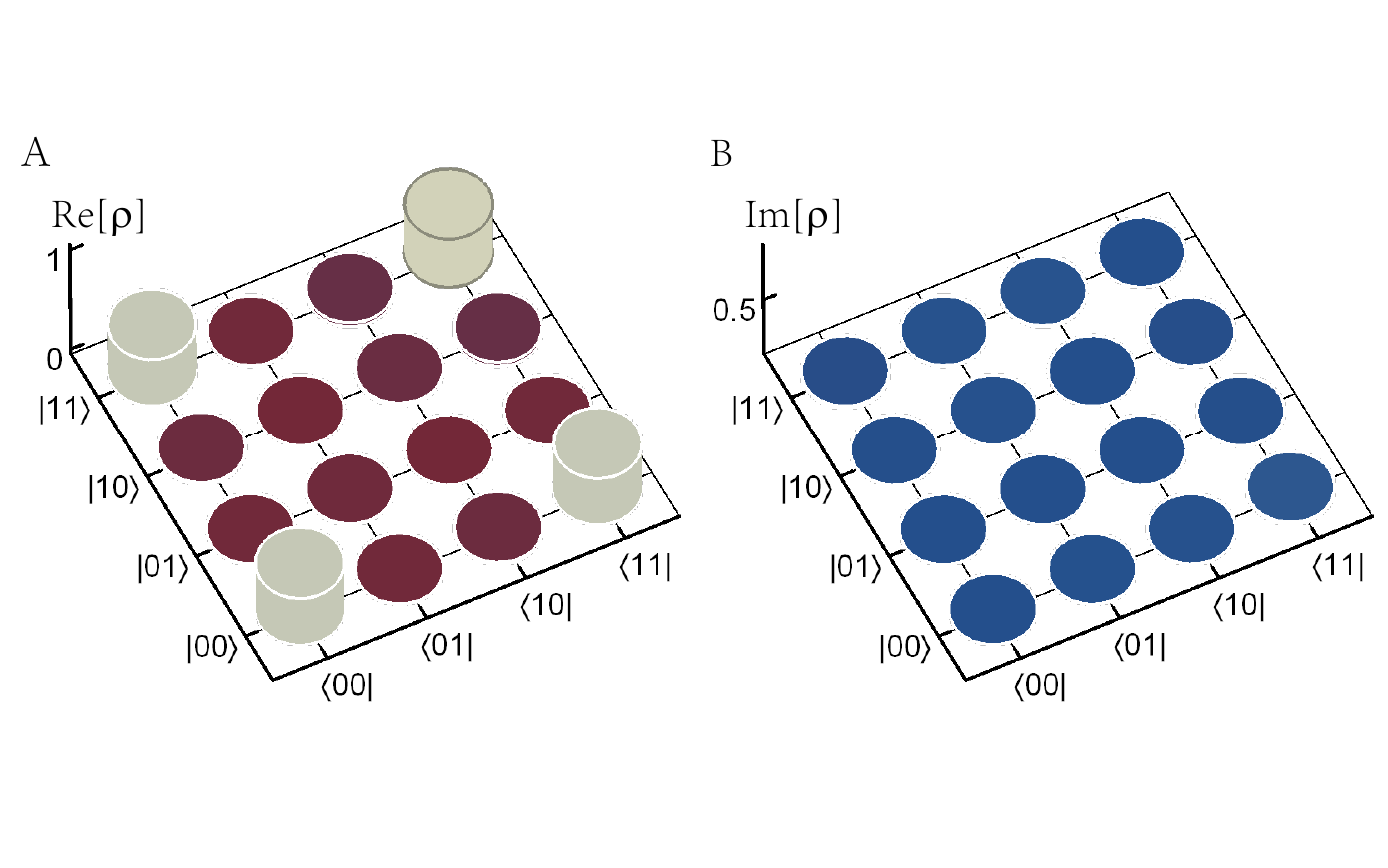}
\caption{{Quantum state tomography.}
The real(A) and imaginary(B) part of density matrix, for the target state $|\psi^+\rangle=\frac{1}{\sqrt{2}}(|00\rangle+|11\rangle)$, is plotted with a fidelity of $0.951\pm0.096$. Over-complete quantum state tomography is used to reconstruct the experimental state, which requires 9 different measurement settings and a total of 36 local projection measurement basis combinations. 
}
\label{FigTomo}
\end{figure}

\section{Experimental setup and demonstration}
\label{Apdx_Werner}

Fig. (3) (in the main text) represents the experimental setup designed on a silicon photonic chip which is used to demonstrate different quantum correlations from the violation of different forms of uncertainty relations~(\ref{LHS_Ent_SM}, \ref{LHS_Steer_SM}, \ref{LHS_BI_2_SM}). Here, a pair of path-encoded entangled photons in the state $  |\psi^{+}\rangle$ ($= (|00\rangle + |11\rangle)/\sqrt{2}$)  are generated via spontaneous four-wave mixing (SFWM) process, by pumping a continuous wavelength laser at 1550.12nm on two spiral waveguides (single photon-pair sources). 
The signal photon (1545.31nm, shown in red) is assumed to be system $A$ (belonging to Alice) while the idler photon (1554.91nm, shown in green) is assumed to be system $B$ (belonging to Bob). The expectation value $\langle\mathcal{A}\,\mathcal{B}\rangle_{|\psi^+\rangle}$,
\begin{eqnarray}
\langle\mathcal{A}\,\mathcal{B}\rangle_{|\psi^+\rangle} = \langle\psi^+|\mathcal{A} \otimes \mathcal{B} | \psi^+\rangle
\label{Expt_psi+}
\end{eqnarray}
 can be experimentally observed from coincidence detection
\begin{eqnarray}
P(a_\mathcal{A},\,b_{\mathcal{B}}) = \frac{C(a_\mathcal{A},\,b_{\mathcal{B}})}{\displaystyle\sum_{\{a,\,b\}=0}^1 C(a_\mathcal{A},\,b_{\mathcal{B}})},
\label{Corr_psi+}
\end{eqnarray}
for the measurement of $\mathcal{A}$ and $\mathcal{B}$ on system $A$ and $B$ prepared in the state $|\psi^+\rangle$, respectively. Here, $C(a_\mathcal{A},\,b_{\mathcal{B}})$ represents coincident counts for the outcome $a$ and $b$ .

In experiment, direct generation of two-qubit Werner state $\rho_W$ is still challenging as it requires the mixture of Bell state and maximal incoherent state, which can be described by 
\begin{eqnarray}
\rho_W &=&p\rho_{|\phi^-\rangle}+(1-p)\,\frac{\rm{I}\otimes \rm{I}}{4} \nonumber \\
&=& \frac{\alpha\, (\rm{I}\otimes \sigma_x) \,\rho_{|\psi^{+}\rangle}\,(\rm{I}\otimes \sigma_x) + \beta\, (\rm{I}\otimes \, \sigma_y) \,\rho_{|\psi^{+}\rangle}\,(\rm{I}\otimes \, \sigma_y) + \gamma  \,\rho_{|\psi^{+}\rangle} + \delta \, (\rm{I}\otimes \sigma_z) \,\rho_{|\psi^{+}\rangle}\,(\rm{I}\otimes \sigma_z)}{\alpha+\beta+\gamma+\delta},
\label{W_State_SM}
\end{eqnarray}
where $\frac{\alpha}{\alpha+\beta+\gamma+\delta}=\frac{1+3p}{4}, \beta=\delta=\gamma=\tfrac{1-p}{4} (\alpha+\beta+\gamma+\delta)$, $\rho_{|\phi^-\rangle}$ and $\rho_{|\psi^{+}\rangle}$ are the density matrix of the states $|\phi^-\rangle$ ($=(|01\rangle-|10\rangle)/\sqrt{2}$) and $|\psi^{+}\rangle$, respectively. The expectation value $\langle\mathcal{A}\,\mathcal{B}\rangle_{\rho_W}$ for the Werner state $\rho_W$ has been experimentally realized from the statistical mixing of identity (\rm{I}) and Pauli rotations ($\sigma_x$,\,$\sigma_y$,\, $\sigma_z$)  on observable $\mathcal{B}$ as given by
\begin{eqnarray}
\langle\mathcal{A}\,\mathcal{B}\rangle_{\rho_w} = \frac{\alpha \langle\mathcal{A}\,\mathcal{B}_x\rangle_{|\psi^+\rangle} + \beta \langle\mathcal{A}\,\mathcal{B}_y\rangle_{|\psi^+\rangle} +  \gamma \langle\mathcal{A}\,\mathcal{B}\rangle_{|\psi^+\rangle} +  \delta \langle\mathcal{A}\,\mathcal{B}_z\rangle_{|\psi^+\rangle}}{\alpha+\beta+\gamma+\delta},
\end{eqnarray}
where $\mathcal{B}_i=\sigma_i\,\mathcal{B}\,\sigma_i$, $i\in\{x,\,y,\,z\}$. The $\langle\mathcal{A}\,\mathcal{B}\rangle_{\rho_W}$ is calculated from the experimentally observed data for $p=0.0,0.2,0.4,0.6,0.8,1.0$, and the corresponding mixing weights $\{\alpha, \, \beta\}$ are 
\begin{eqnarray}
p&&=0.0\rightarrow \{\alpha=5,\beta=5\}, \hspace{1cm} p=0.6\rightarrow \{\alpha=14,\beta=2\}, \nonumber \\
p&&=0.2\rightarrow \{\alpha=8,\beta=4\}, \hspace{1cm} p=0.8\rightarrow \{\alpha=17,\beta=1\}, \\
p&&=0.4\rightarrow \{\alpha=11,\beta=3\}, \hspace{1cm} p=1.0\rightarrow \{\alpha=20,\beta=0\}. \nonumber
\end{eqnarray}
For example, $p=0.0$ is calculated from the 5 sets of data of each $\langle A\mathcal{B}_i\rangle_{|\psi^+\rangle}$.
In the experiment, the Bell state $|\psi^{+}\rangle$ has been generated with fidelity $0.951\pm0.096$, and the details of quantum state tomography (QST) is shown in the Fig.\ref{FigTomo}. The error bars in terms of standard deviation have been calculated from the 20 sets of data, which result in the order of $10^{-2}$ for entanglement, steering and Bell nonlocality.

\end{document}